\documentclass[11pt]{article}
\pdfoutput=1
\usepackage{fancyhdr}
\usepackage[us,12hr]{datetime}
\usepackage{jcappub}
\usepackage[T1]{fontenc}
\usepackage{fullpage}
\usepackage{amsmath}
\usepackage{graphicx}
\usepackage{amssymb}
\usepackage{mathrsfs}
\usepackage{marvosym}
\usepackage{booktabs}
\usepackage[english]{babel}
\usepackage{amsmath,amssymb,amsbsy,amstext, amsthm, simplewick}
\usepackage{graphicx}
\usepackage[framemethod=tikz]{mdframed}
\usepackage{amsfonts}
\usepackage{framed}
\usepackage{amssymb}
\usepackage{colortbl}
\usepackage{xcolor}
\definecolor{green2}{cmyk}{0, 1, 0.5, 0}
\definecolor{lightgreen}{cmyk}{0.2, 0, 0.2, 0.2}
\definecolor{lightgray}{cmyk}{0.1,0.2,0,0.1}
\definecolor{lightgray2}{cmyk}{0.4,0.4,0,0.8}
\definecolor{black}{cmyk}{1.0,1.0,1.0,1.0}
\definecolor{brown}{rgb}{0.79, 0.16, 0.16}
\definecolor{cerulean}{rgb}{0.0, 0.48, 0.65}
\definecolor{blue2}{rgb}{0.0, 0.2, 0.4}
\definecolor{pc1}{rgb}{0.69, 0.25, 0.21}
\definecolor{royalazure}{rgb}{0.4, 0.2, 0.92}
\definecolor{darkcerulean}{rgb}{0.03, 0.27, 0.89}
\definecolor{magenta1}{rgb}{0.89, 0.08, 0.48}
\definecolor{mycolor}{rgb}{0.122, 0.435, 0.698}
\newmdenv[innerlinewidth=0.5pt, roundcorner=4pt,linecolor=mycolor,innerleftmargin=6pt,
innerrightmargin=6pt,innertopmargin=6pt,innerbottommargin=6pt]{mybox}
\usepackage[skins,theorems]{tcolorbox}
\tcbset{highlight math style={enhanced,
  colframe=mycolor,colback=white,arc=0pt,boxrule=1pt}}

\usepackage[left=28mm,top=24mm,right=28mm,bottom=26mm]{geometry}

\newcommand{\vn}{\mathbf{n}}

\newcommand{\hatbmu}{\widehat{\boldsymbol{\mu}}}
\newcommand{\hatbnu}{\widehat{\boldsymbol{\nu}}}
\def\sun{{\sf SU}($N$)~}
\def\un{{\sf U}($N$)~}
\def\sunl{{\sf SU(N)}~}
\def\unl{{\sf U(N)}~}

\begin{document}
\begin{titlepage}
\setcounter{page}{1} \baselineskip=15.5pt \thispagestyle{empty}
\bigskip\
\vspace{1cm}
\begin{center}
\textcolor{brown}{\textbf{\textsc{\fontsize{20}{48} \bfseries Finite $N$ unitary matrix model}}}
\end{center}
\vspace{4mm}
\begin{center}
\Large{\textcolor{black}{Raghav G. Jha}}


\vskip 1pt
\textit{\fontsize{13}{26}\selectfont  Perimeter Institute for Theoretical Physics, 
Waterloo, Ontario N2L 2Y5, Canada}
\vskip 3pt
\fontsize{13}{26}\texttt{\Email ~ 
\href{mailto:rjha1@perimeterinstitute.ca}{rjha1@perimeterinstitute.ca}}

\end{center}
\vspace{1.2cm}
\hrule \vspace{0.3cm}
\noindent {\textsc{\bfseries Abstract:}} We consider one-plaquette unitary matrix model at 
finite $N$ using exact expression of the partition function for both \sun 
and \un groups.
\\[0.1cm]
\vskip 10pt
\hrule

\tableofcontents 
\end{titlepage} 

\section{Introduction}
The study of matrix models has been important in understanding
and gaining deep insights into the structure of gauge theories. 
Their range of applicability is wide and some examples include -- 
the geometry of random surfaces, two-dimensional quantum gravity, 
statistical mechanics of spin systems. They also have connections to 
supersymmetric gauge theories and non-perturbative formulations of 
superstring theory.  A class of matrix models has 
also been studied in the context of exhibiting properties similar to the
four-dimensional theory of strong interactions (QCD). 
The techniques of strong coupling expansion and lattice methods 
\cite{Wilson:1974sk} along with approximate recursion relations 
\cite{Migdal:1975zg} were pioneered to understand the 
properties of QCD. Around the same time, 
another important direction was pursued by 't~Hooft in finding an 
expansion parameter, independent of the scale to 
study QCD by considering 1/$N$ expansion
\cite{tHooft:1973alw}. In this setting, 
the Feynman diagrams drawn in the double-line notation can be 
organized to follow an expansion in powers of $N^{\chi=2-2g}$
and were argued to be analogous to the topological expansion in string theory. 
The large $N$ limit was then 
used to solve the two-dimensional model of QCD in the light-cone gauge 
which is now also known as 't~Hooft model \cite{tHooft:1974pnl}. 
This model has many similarities with the four-dimensional theory but is 
exactly solvable in the planar limit. In higher dimensions, 
the success has been scarce but it is at least known that the large $N$ limit 
does not alter a distinctive feature of QCD -- ultraviolet freedom, 
which is obvious from the $N \to \infty$ limit of the two-loop QCD $\beta$-function 
which has the leading coefficient with the correct sign. Following these developments, 
the large $N$ limit of gauge theories have been extensively explored
for a wide class of gauge theories. In this 
topological limit, a remarkable simplification occurs, where only the 
planar diagrams survive, when we keep $\lambda = g_{\text{YM}}^2 N$ fixed 
and take $ N \to \infty$.  
The interplay between the large $N$ limit of gauge theories and string theory 
has significantly improved our understanding of strongly coupled gauge 
theories and resulted in the AdS/CFT conjecture. In this work, we 
consider a unitary matrix model that is equivalent to two-dimensional pure 
QCD. The two-dimensional models such as this
have rich behavior but are trivial since gauge field is non-dynamical 
(absence of degrees of freedom) and the expectation value of observables 
can be calculated exactly as done in \cite{Wadia:2012fr, Gross:1980he, 
Goldschmidt:1979hq}. In this model, the Wilson loop obeys 
\textit{area law} for all couplings since the Coulomb potential is always 
linear and exhibits both ultraviolet freedom and infrared 
prison. The two-dimensional lattice gauge theories with Wilson action were shown to exhibit third-order phase transition at large $N$. This transition came as a surprise when it was first observed but now we have a better understanding through AdS/CFT correspondence that these phase transitions at large $N$ and finite volume occurs in several holographic models and are related to transitions between different black hole solutions in supergravity \cite{Witten:1998zw, Catterall:2017lub}.
The occurrence of this phase transition in the large $N$ limit signifies that non-analytic behaviour can occur even in the simplest of models which makes it clear that the strong coupling expansions cannot be usually smoothly continued to 
the limit of weak coupling. This two-dimensional lattice gauge theory with \sun  or \un  gauge group is often described in terms of single plaquette action over compact unitary group manifold and will hereafter be referred to as the GWW model 
after the authors of \cite{Gross:1980he,Wadia:2012fr}. 
The unitary matrix models of this kind have also been studied 
from the holographic point of view 
\cite{Kabat:1999hp,Kabat:2001ve}. 
The interest in this model stems from two main reasons, 1) This model admits an exact 
solution for any $N$ and coupling, 
2) This model is closely related to 
one of the simplest strings (minimal superstring theory) with manageable non-perturbative 
description (Type 0 strings in one dimension) \cite{Klebanov:2003wg}.  
In fact, it was shown that unitary matrix models in the double scaling limit 
($N \to \infty$ \textit{\&} $\lambda \to  \lambda_{\text{critical}}$ with 
some well-defined relation between $N$ and $|\lambda - \lambda_{\text{critical}}|$)
is described by the Painlev\'{e} II equation \cite{Periwal:1990gf}.  

The GWW model and several of its 
modifications have been well-studied 
using various techniques in the $N \to \infty$ limit. 
The finite $N$ limit has been surprisingly less explored 
but recently attracted some attention \cite{Okuyama:2017pil}. 
Even in these explorations, the gauge group considered
was \un  since it is a little easier to handle and in the planar
limit, which is mostly of interest, there is no distinction with \sun . 
However, at finite $N$, they have \emph{different} behaviour and independent studies of 
both $\mathbb{G}= $ \sun  and \un  gauge groups are desirable and to 
our knowledge no treatment of this unitary matrix model at finite 
$N$ for \sun  gauge group has been done yet. 
This paper aims to fill this gap in the literature. We derive an
expression for the partition function considering \sun 
gauge group and study observables in the finite $N$ limit. 
In case the qualitative behaviour is not severely altered by 
addition of matter fields i.e. presuming this phase transition 
is not turned into a smooth crossover, then these studies may be useful
in understanding black hole solutions and stringy corrections
by considering the finite $N$ limit of matrix models \cite{Susskind:1997dr} 
since the transition at $\lambda_{\text{critical}} = 2$ is supposed to separate 
the supergravity regime from the perturbative gauge theory regime. There 
are only a few matrix models where the finite 
$N$ regime is exactly solvable for any coupling and this is one such model. 
In this respect and various others, a finite $N$ study of these class 
of matrix models would be useful in probing the quantum effects of gravity using holography. 

We briefly outline the structure of the paper. In Section \ref{sec:2}, we write down the exact expression 
of the partition function for \emph{special unitary matrix} model 
at any coupling and $N$ in terms of determinant of a Toeplitz 
matrix\footnote{Named after 
German mathematician Otto Toeplitz (1881 - 1940), each descending diagonal from left to right 
in this matrix is constant and matrix element $(j,k)$ depends on $j-k$} and sketch the well-known phase structure of this model at 
large $N$. In Section \ref{sec:3}, we numerically calculate the corrections to the planar result 
for free energy in both the phases and 
show that it is simpler to deduce the instanton corrections in the strong coupling phase 
since the $1/N$ (or genus) expansion terminates at genus-zero for \un  unitary model. 
However, this is not the case for the \sun  model which has 
contributions to the free energy
at small $N$ in both the phases. We also provide results for the 
Wilson loop in the 
\sun  matrix model using our exact expression and show that in the large 
$N$ limit, they converge to the known results for the \un  unitary model. 
We end with a brief mention of some open questions which can be addressed 
in the context of \sun  one-plaquette matrix model and other 
related unitary matrix models. 

\section{Partition function and observables} 
\label{sec:2} 
The central object in the study of matrix models is the partition function which is determined by integration 
over some compact group manifold. However, there
exists only a handful of models which can be solved 
exactly \cite{Kazakov:2000aq} and most of these proceeds 
by reduction of $N^{2}$ matrix degrees of freedom to $N$ by exploiting some symmetry in the large $N$ limit. 
The simplest among all such matrix models is the well-studied one-matrix model. 
For a nice review for unitary matrix models, we refer the reader to \cite{Rossi:1996hs}. The lattice 
Wilson action in two dimensions is equivalent to one-plaquette model by considering the integration 
over the unitary group. This has also been studied using character expansion\footnote{It is useful to sometimes think of  character expansion as Fourier expansion on a compact group manifold}
which is discussed in \cite{Balantekin:2000vn,Bazavov:2019qih}.
It was shown \cite{Migdal:1975zg} that in two dimensions, 
the expansion in terms of characters constitute the recursion relations
through which one can exactly solve the model.
For the large $N$ analysis, the saddle point methods 
\cite{1982NuPhB.194..107G} are often used 
as was done in \cite{Gross:1980he}. However, the saddle point methods are not useful in 
extracting sub-leading orders in the 1/$N$ expansion for which the method of orthogonal 
polynomials is usually used as was done in \cite{Goldschmidt:1979hq}. 
The general partition function can be schematically written as:
\begin{equation}
Z = \frac{1}{\text{Vol}(\mathbb{G})} \int \prod_{\text{links} ~ l} \mathscr{D}U_{l} \prod_{\Box} Z_{\Box},
\end{equation}
where $\Box$ denotes the plaquette on whose perimeter we take the 
ordered product of the 
link matrices $U$ of size $N \times N$. In order to compute expectation 
values in the two-dimensional model, 
one has to make a choice between one-plaquette and heat kernel actions. 
The partition function of the two-dimensional Yang-Mills 
theory based on the heat kernel action is 
written in terms of sum over all irreducible representations of the unitary 
gauge group. We will use the one-plaquette action in this work similar to
\cite{Wadia:2012fr, Gross:1980he} which can be expressed as\footnote 
{Some authors use $g$ to denote $1/\lambda$ or $2/\lambda$. This distinction is 
clear from the coupling where phase transition occurs.}:
\begin{equation}
\label{eq:main111} 
S(U) = \frac{N}{\lambda} \Big[ \mathrm{Tr} 
\Big( \prod_{\Box} U \Big) + \mathrm{Tr} \Big( \prod_{\Box} U^{\dagger} \Big) \Big], 
\end{equation}
where $\prod_{\Box} U$ is the product of links around a 
plaquette and shorthand for 
$U_{\mu}(\textbf{n}) U_{\nu}(\vn+\hatbmu) 
U^{\dagger}_{\mu}(\vn+\hatbmu + \hatbnu)U^{\dagger}_{\nu}(\vn+\hatbnu)$. 
The convention used is such that $U_{\mu}(\textbf{n})$ denotes a link starting from site $\textbf{n}$ and going in $\mu$-direction. 

It was found that for this model in the 
$ N \to \infty$ limit with fixed $\lambda = g_{\text{YM}}^2 N$, 
one observes a discontinuity in the third derivative 
of the free energy corresponding to a third-order phase transition. 
This means that one loses the analytic structure for 
even simple actions in the large $N$ limit and the continuation 
from the weak coupling to strong coupling is non-trivial. 
This transition is not like the usual phase transitions in statistical mechanics 
which occur in the infinite volume limit. Here, it occurs in a finite volume 
(single plaquette) for an infinite rank gauge group.   

We denote the full partition function of the model with $\mathcal{Z}$ 
and since in two dimensions, we can treat all the 
plaquettes as independent, we deal with just a single plaquette
\cite{Drouffe:1978dn, Gross:1980he}, which
we will refer to as one-plaquette partition function,  
$ Z = \mathcal{Z}^{1/N_{s}N_{t}}$, where
$N_{s}N_{t}$ are the number of nodes in the lattice and equals the 
number of plaquettes. 
It is given by: 
\begin{equation}
\label{eq:main model}
Z = \int_{\substack{\unl \\ \text{or~} \sunl}}
\mathscr{D}U \exp \Bigg [\frac{N}{\lambda} \mathrm{Tr} \Big(U^{\dagger}  + U \Big) \Bigg].
\end{equation}
It is known that the partition function
for the \un  matrix model given above 
can be written in terms of Toeplitz determinant\footnote{The determinant of 
infinite size Toeplitz matrix has a well-defined 
asymptotic behaviour limit due to a theorem by Szeg\"{o}.}  
given by \cite{Bars:1979xb, Goldschmidt:1979hq}:
\begin{equation}
\label{eq:UN1}
Z (N, \lambda) = \text{Det}\Bigg[I_{j-k}\Big(\frac{2N}{\lambda}\Big)\Bigg]_{j,k = 1 \cdots N},
\end{equation}
where $I_{\nu}(x)$ is the modified Bessel function of first kind or Bessel function of imaginary argument of 
order $\nu$. Note that the argument of the Bessel function is twice the prefactor of action in (\ref{eq:main model}). 
The appearance of partition function in terms of determinant has deep connections to the 
notion of integrability and special differential equations. 
For instance, the determinant of Toeplitz matrices also play a role 
in the context of the Ising model, for a partial set of references, see 
\cite{deift2012toeplitz,1966PhRv..149..380W}. 
Instead of working with the partition function given in (\ref{eq:main model}), one can 
also consider more general unitary matrix model with source arbitrary 
$N \times N$ matrix $A$ as 
was considered in \cite{Brezin:1980rk}. 
The action for this model is given by:
\begin{equation}
\label{eq:source1} 
Z = \int \mathscr{D}U \exp \mathrm{Tr} \Big[UA^{\dagger}  + AU^{\dagger}\Big], 
\end{equation}
where $U$ is product of four ordered links. The exact form of $Z$ was derived in the large 
$N$ limit and it was shown that the strong and weak coupling regimes in this model are characterized by 
$ (1/N) \mathrm{Tr}(A^{\dagger}A)^{-1/2}$. One gets back the usual GWW model by setting $ 
A = \mathbb{I}/\lambda$. The model in (\ref{eq:source1}) was related to a specific 
$N \times N$ Hermitian matrix model following some parameter tuning 
in \cite{Brezin:2010ar} and an exact solution was found. In this work we will 
only consider (\ref{eq:main model}) and derive the exact expression for the partition function 
when the integration is over \sun  group in (\ref{eq:main model}). 

The analysis for \sun  is similar to the one for \un  
except that now we have the constraint that 
the product of eigenvalues should satisfy $\prod_{j=1}^{N} e^{i\alpha_{j}} = 1$. We start with the 
one-plaquette partition function: 
\begin{equation}
Z = \int_{SU(N)} \mathscr{D}U \exp \Bigg[ \frac{N}{\lambda}  \Bigg( \mathrm{Tr}  \prod_{\text{single~}\Box} U  + 
\mathrm{Tr}  \prod_{\text{single~}\Box }  U^{\dagger} \Bigg) \Bigg],
\label{eq: Z single} 
\end{equation}
where the measure is given by: 
\begin{equation}
\label{eq:measure1}
\mathscr{D}U = 
\frac{1}{N!} \prod_{j=1}^{N} 
\underbrace{\sum_q \delta \Big(\sum_{m=1}^{N} 
\alpha_{m} - 2q\pi \Big)}_{\text{\sun  constraint}} \frac{d\alpha_{j}}{2\pi} 
\prod_{j<k} \Big(e^{i\alpha_{j}} - e^{i\alpha_{k}}\Big) \Big(e^{-i\alpha_{j}} - e^{-i\alpha_{k}}\Big),
\end{equation}
and the constraint can be written as:
\begin{equation}
\frac{1}{2\pi} \sum_{p=-\infty}^{\infty} e^{ip \sum \alpha_{m}}  ~~ .  
\end{equation}
By using the representation of the modified Bessel function as:
\begin{equation}
\label{eq:def_I}
I_{k-j-p}\Big(\frac{2N}{\lambda}\Big) = I_{j-k+p}\Big(\frac{2N}{\lambda}\Big) = 
\frac{1}{2\pi} \int_{0}^{2\pi} ~ e^{\frac{2N}{\lambda} \cos \alpha} e^{i(j-k+p)\alpha} d\alpha, 
\end{equation}
where, $e^{i\alpha}$ are the eigenvalues of $U$ and 
(\ref{eq:def_I}) fills the corresponding 
$j^{\text{th}}$ and $k^{\text{th}}$ element of the matrix  
$\mathcal{M}_{\alpha} = I_{j-k+\alpha} (\frac{2N}{\lambda})$ 
and using (\ref{eq: Z single}) through (\ref{eq:def_I}), we obtain the partition function for \sun  matrix model as:
\begin{equation}
\label{eq:5} 
\tcbhighmath[drop fuzzy shadow]{Z (N,\lambda) = \sum_{p=-\infty}^{\infty} 
\text{Det}\Bigg[I_{j-k+p}\Big(\frac{2N}{\lambda}\Big)\Bigg]_{j,k = 1 \cdots N}.}
\end{equation}
In contrast to the expression for the exact partition function 
for \un  matrix model i.e. (\ref{eq:UN1}), 
there is an additional sum over the index $p$ from the 
constraint in (\ref{eq:measure1}).
In this paper, we study different observables using this 
partition function. However, for practical purposes of 
computation using Mathematica, we simply replace the 
$\infty$ in (\ref{eq:5}) by a large number. We checked that $\left|p \right| \le 15 $ suffices i.e., 
\begin{equation}
\label{eq:main1} 
Z (N, \lambda) = \underbrace{\sum_{p=-15, \neq 0}^{15} \text{Det}
\Bigg[I_{j-k+p}\Big(\frac{2N}{\lambda}\Big)\Bigg]
}_{\text{due to \sun }} 
+ \text{Det}\Bigg[I_{j-k}\Big(\frac{2N}{\lambda}\Big)\Bigg]. 
\end{equation}
This partition function enables us to evaluate free energy which can be 
written as a sum over genus expansion for this model: 
\begin{equation}
\label{eq:generalF} 
F(N,\lambda) = \sum_{g \in \mathbb{N}} F_{g} N^{2-2g}  + \mathcal{O}(e^{-N}), 
\end{equation}
where $\mathbb{N}$ denotes the set of non-negative integers. 
The exact result for the leading coefficient, $F_{0}$, in the planar limit 
is given by:
\begin{equation}
\label{eq:exprF0} 
F_{0} =  \hspace{2mm} \begin{cases}
    \frac{-1}{\lambda^2} \hspace{38mm}    \text{$\lambda \ge 2$}   \\
     \frac{-2}{\lambda}  - \frac{1}{2} \ln \frac{\lambda}{2} + \frac{3}{4} 
     \hspace{17mm} \text{$\lambda < 2$}
      \end{cases}.
    \end{equation} 
Another important observable in unitary matrix models and the one we 
consider are the Wilson loops. The finite $N$ analysis of Wilson loops in 
various representations for 
\un  gauge theory was recently done in \cite{Okuyama:2017pil}. 
An expression for the 
expectation value of normalized winding Wilson loops
defined as:
\begin{equation}
\mathcal{W}_{k}(N,\lambda) = \frac{1}{N} \Big \langle \mathrm{Tr} 
\Big(\prod_{\mathcal{C}} U\Big)^{k} \Big \rangle, 
\end{equation}
was given in terms of  $ \mathrm{Tr}(\mathcal{M}_{k}/\mathcal{M}_0)$ 
with $k \in \mathbb{Z}^{+}$ denoting 
the winding number and the expectation value is computed over a 
closed contour $\mathcal{C}$.  
A similar expression for the \sun  case is yet unknown. 
Note that like the partition function, we can write $\mathcal{W}_{k, \mathcal{C}}(N,\lambda)$
as $W_{k}(N,\lambda)^{N_{s}N_{t}}$, where the contour is of time 
extent $aN_{t}$ and spatial extent $aN_{s}$. 
The single winding ($k=1$) Wilson loop is related to the derivative of the 
free energy and is given by:
\begin{equation}
W_{1}(N,\lambda)  =  \frac{-\lambda^2}{2N^2} \frac{\partial \ln Z}{\partial \lambda} =  
\hspace{2mm} \begin{cases}
    \frac{1}{\lambda} \hspace{17mm}    \text{$\lambda \ge 2$}   \\
     1 - \frac{\lambda}{4} \hspace{10mm} \text{$\lambda \le 2$}
      \end{cases}.
    \end{equation} 
For this \un  matrix model as mentioned above, there is another equivalent definition of $W_{1}$ given by:  
\begin{equation}
W_{1}(N,\lambda) = \mathrm{Tr} ~ \Big(\frac{\mathcal{M}_{1}}{\mathcal{M}_{0}}\Big),
\end{equation}
where $\mathcal{M}_{\alpha}$ is defined below (\ref{eq:def_I}). 
We will present results for the free energy and Wilson loops 
in Section \ref{sec:3} for the \sun  and \un  models.  

Even though this matrix model is one of the simplest it has a wide range 
of interesting features. In \cite{Marino:2008ya}, by 
considering the trans-series solutions of the pre-string equation to 
obtain instanton corrections, it was deduced that the instanton action
vanishes at the critical point i.e. $\lambda = 2$  and it was concluded that the GWW transition is caused by the effect of instantons\footnote{Usually, when one thinks of large $N$ limit, it seems that the instanton corrections which goes 
as $\exp(-A/g_{s}) \sim  \exp(-AN/\lambda)$ are insignificant 
with $A$ denoting the instanton action. 
In fact, a more general form is $\exp(- F(\lambda)N)$, where 
$F$ is some non-negative function of the coupling $\lambda$ 
and proportional to $A$. When $F(\lambda)=0$ corresponding to  
vanishing action, the 
exponentially suppressed instanton contribution to $1/N$ 
expansion becomes important. It turns out that for the 
GWW model this happens
exactly at $\lambda_{\text{critical}} =2$ where the third-order 
phase transition takes place. Therefore, one also refers to 
these as instanton driven large $N$ phase transitions and physically 
relate them to condensation of instantons. We thank 
M. Mari\~{n}o for email correspondence regarding the one-plaquette model 
and his book ~`Instantons and Large $N$' for clarification regarding the instanton contributions}. 
There exist other examples where a similar 
phenomenon occurs \cite{Gross:1994mr}. The behaviour of the corrections to the 
planar result due to $1/N$ and instanton have 
been well-studied. This model also exhibits 
resurgence behaviour thoroughly explored in \cite{Ahmed:2017lhl}. 
One of the striking features of these studies, which is clearly evident
in our results is that the strong coupling phase has no $1/N$ corrections 
\cite{Goldschmidt:1979hq} but only $\mathcal{O}(e^{-N})$ 
corrections from the contributions due to instantons. In 
GWW model, both gapped and ungapped phases have instanton 
corrections, albeit of different nature. In the ungapped phase, 
the eigenvalues of the holonomy matrix fill the circle while in 
the gapped phase they are distributed over some interval. When the 
eigenvalue distribution is restricted to some domain, 
it is called a one-cut (or single interval) distribution/solution. 
In the ungapped phase, the instanton contribution to the free energy 
can simply be evaluated by subtracting genus-zero contribution from 
the total free energy. As we will see, this does not hold for the \sun 
 model. The distribution of eigenvalues, which is one of the central 
 objects in these matrix models, have been studied in both the phases 
 for the \un  model and is given by: 
\begin{equation}
\rho(\lambda,\theta) =  \hspace{2mm} \begin{cases}
    \frac{2}{\pi \lambda} \cos\Big(\frac{\theta}{2}\Big) \sqrt{\frac{\lambda}{2} - \sin^{2} \frac{\theta}{2}}\hspace{17mm}    
    \text{$\lambda < 2$}   \\
     \frac{1}{2\pi}\Big(1 + \frac{2}{\lambda} \cos \theta \Big) \hspace{29mm} \text{$\lambda > 2$}
      \end{cases}.
    \end{equation} 
In the gapped phase, the distribution is only supported on some 
interval $ \theta \in [-a,a]$
while it is uniform in the ungapped phase. For \sun  unitary 
matrix model, there are corrections to the distributions above which was discussed in 
\cite{Campostrini:1994ih} and will not be further discussed in this work. 

\begin{figure}[htbp] 
\centering 
\includegraphics[width=0.75\textwidth]{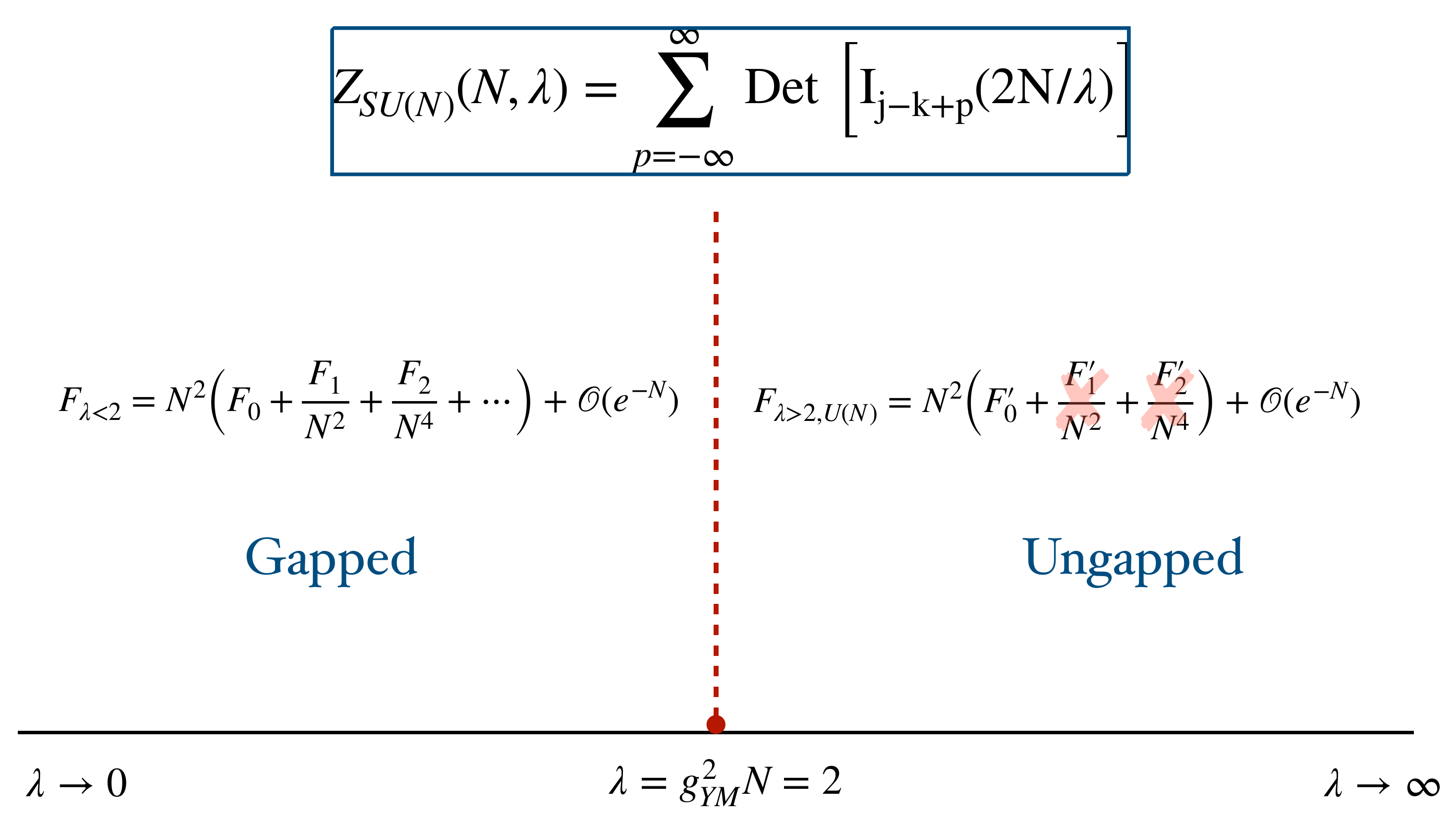} 
\caption{\label{fig:plot1}
A telegraphic summary describing the phase diagram of the one-plaquette 
matrix model at large $N$ and the 
exact expression of the partition function for \sun  model as 
given in the main text.}
\end{figure}

\section{Results \textit{\&} Conclusions} 
\label{sec:3} 
In this section we present the results obtained using (\ref{eq:main1}) and 
(\ref{eq:UN1}) for \sun and \un groups respectively.   
We primarily focus on the free energy for different couplings to emphasize 
behaviours in both phases and at the critical coupling. 
Our results converge to expected results in (\ref{eq:exprF0}) when we take large $N$, 
while also probing the finite $N$ 
coefficients i.e. $F_{g}$ with $g \neq 0$ according to (\ref{eq:generalF}). 
In the weak coupling limit, $\lambda < 2$, the free energy up to 
genus-two was calculated in \cite{Goldschmidt:1979hq} and given by: 
\begin{small} 
\begin{align}
\label{eq:Gold1} 
F(\lambda, N) =  F_{0} -\frac{1}{N^2}\Big(\frac{1}{12} - \ln A  - 
\frac{1}{12} \ln N - \frac{1}{8} \ln 
(1 - (\lambda/2))\Big)  -\frac{1}{N^4}\Big( \frac{3\lambda^3}{1024} 
\frac{1}{(1 - (\lambda/2))^{3}} + \frac{1}{240}\Big), 
\end{align}
\end{small} 
where, $ A = 2^{\frac{7}{36}} \pi^{\frac{-1}{6}} \exp\Big[\frac{1}{3} + 
\frac{2}{3} \int_{0}^{\frac{1}{2}} \ln \Gamma(1+x) dx \Big] = 
1.2824271291 \cdots$ is the Glaisher-Kinkelin constant and is 
related to the derivative of zeta function as, 
$ \zeta^{\prime}(-1) =  \frac{1}{12} - \ln A$. 
A general expression for genus $g$ free energy $F_{g}(\lambda)$ is also known 
\cite{Goldschmidt:1979hq,Periwal:1990gf,Marino:2008ya, Okuyama:2017pil} and 
can be written as:
\begin{equation}
F_{g}(\lambda) = \frac{B_{2g}}{2g(2g-2)} + \frac{1}{\Big(\frac{2}{\lambda} -1\Big)^{3g-3}} 
\sum_{n=0}^{g-2} C_{n}^{g} \lambda^{-n}, 
\end{equation}
where $B_{2g}$ is the Bernoulli number\footnote{We
note that if we calculate the orbifold Euler characteristic of 
the moduli space of Riemann surfaces of genus $g$ with $n$ marked points 
i.e. $ \chi(\mathcal{M}_{g,n})$ using Harer-Zagier formula and set $n=0$ we also 
obtain $\frac{B_{2g}}{2g(2g-2)}$}. 

\begin{figure}[htbp] 
\centering 
\includegraphics[width=0.75\textwidth]{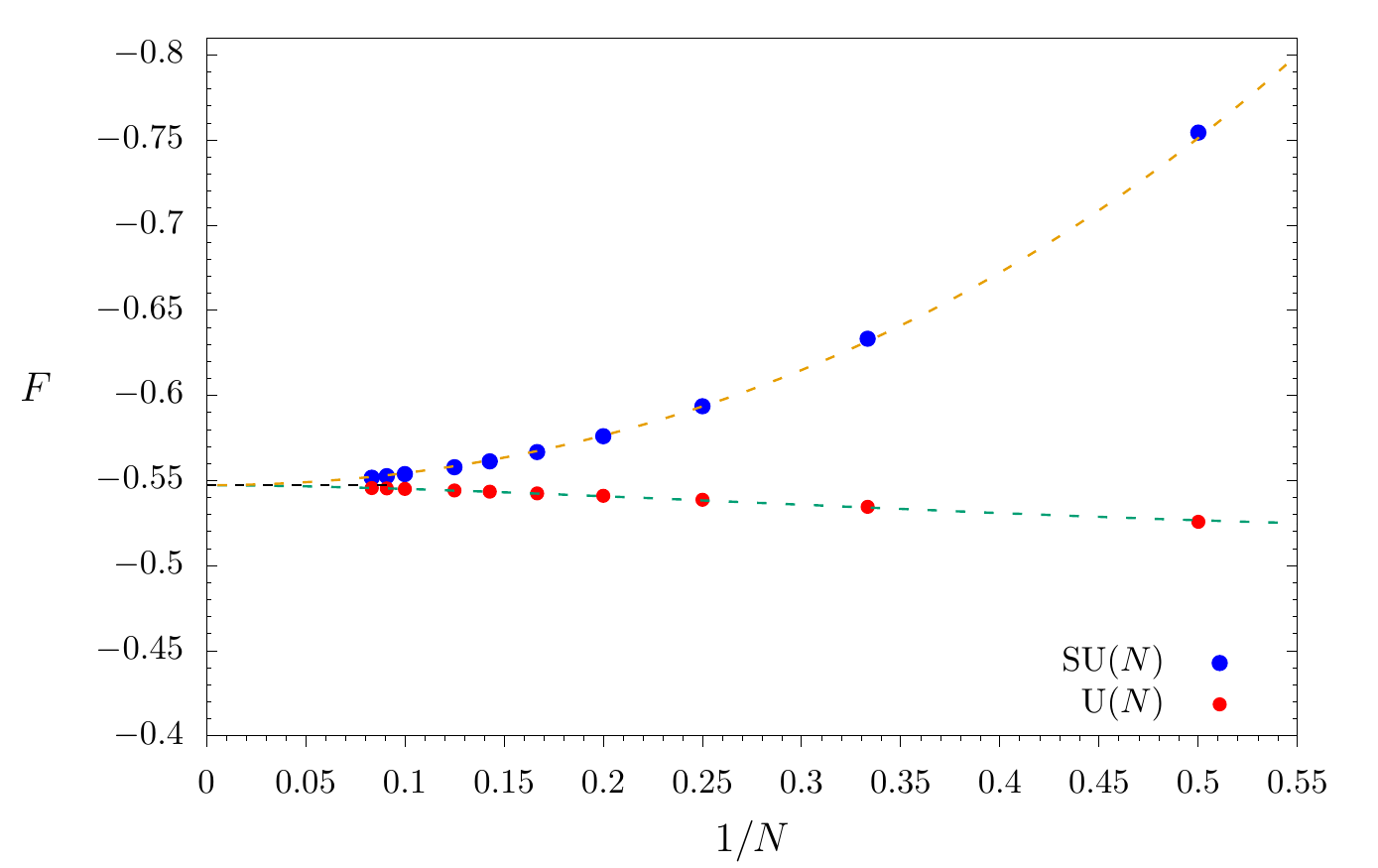}
\caption{\label{fig:plot2}The free energy (normalized by $N^2$) for $\lambda = 4/3$ 
plotted against 
1/$N$. The dashed green line is (\ref{eq:Gold1}) from \cite{Goldschmidt:1979hq} 
and the 
dashed black line is the $N \to \infty$ value. The analytic
 expression corresponding
to the dashed yellow line is yet unknown.}
\end{figure}

A general expression for the \sun  case is unknown but is 
expected to have some similarity since the origin of these Bernoulli numbers 
are in the volume of the \un  group which is similar to that of 
\sun. In Figure \ref{fig:plot2}, we compute the free energy for
$\lambda=4/3$ and show the results for \sun  and \un  models.

\begin{figure}[htbp] 
\centering 
\includegraphics[width=0.75\textwidth]{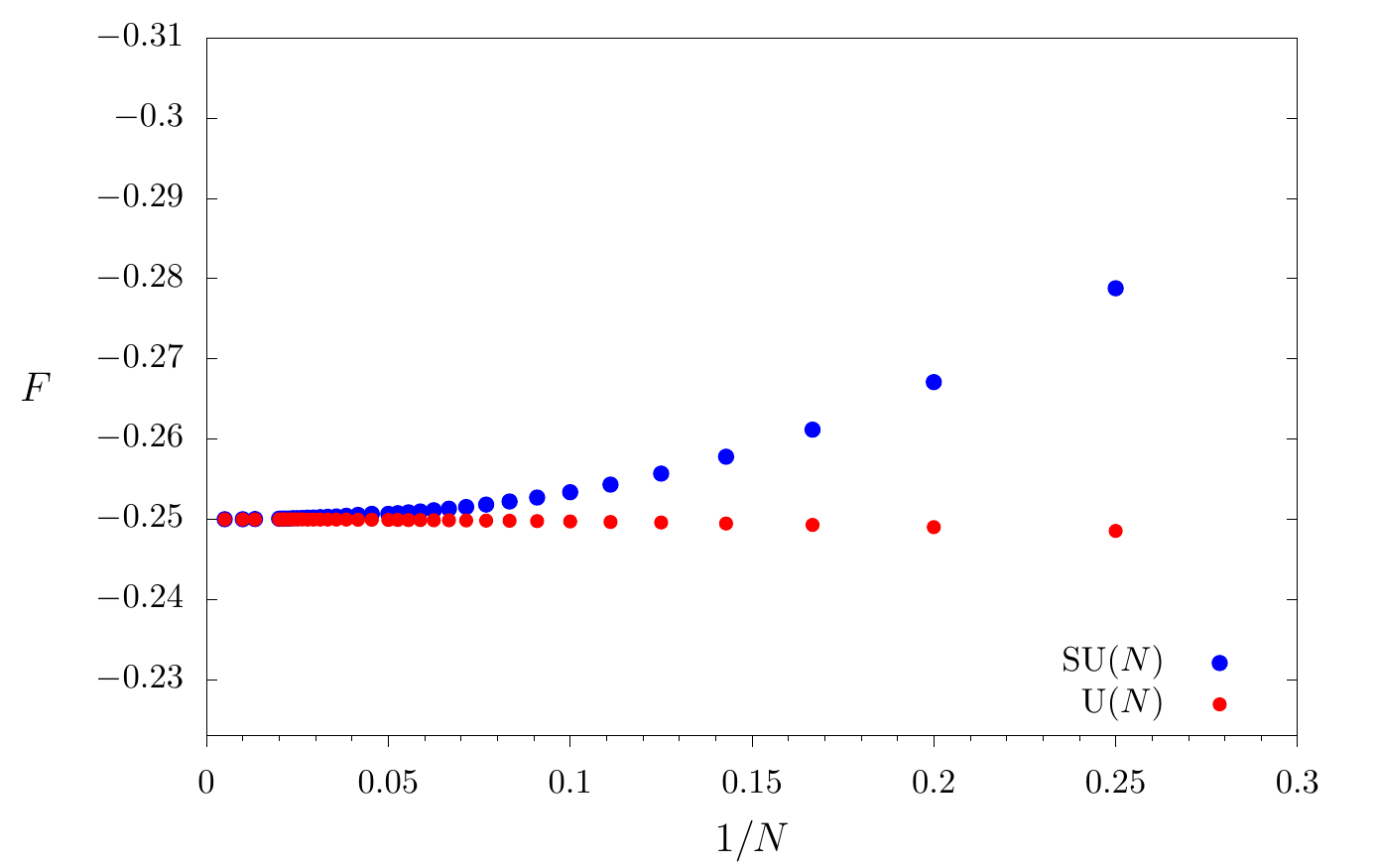}
\caption{\label{fig:plot3}The dependence of the free energy (normalized by $N^2$) 
for \sun  and \un  model at $\lambda=2$ on $N$. 
In the $N \to \infty$ limit, the exact value is $F_{0} = -0.25$.}
\end{figure}

We then consider $\lambda_{\text{critical}} = 2$ and plot the 
results in Figure \ref{fig:plot3}. The free energy has noticeable difference
for \sun  and \un  groups for $N < 15$ (which is not the case for other couplings) 
and we have explored up to $N=100$ for this case. One plausible explanation 
for this behaviour is that at the critical coupling the instanton contributions are more 
important compared to any other $\lambda$ (for fixed $N$) and the difference
between \sun and \un instanton sectors are therefore most significant. It might be possible to 
explain this by a systematic finite $N$ instanton contributions with \sun  group
and comparing with the known results in the \un  models \cite{Marino:2008ya}. 

\begin{figure}[htbp] 
\centering 
\includegraphics[width=0.75\textwidth]{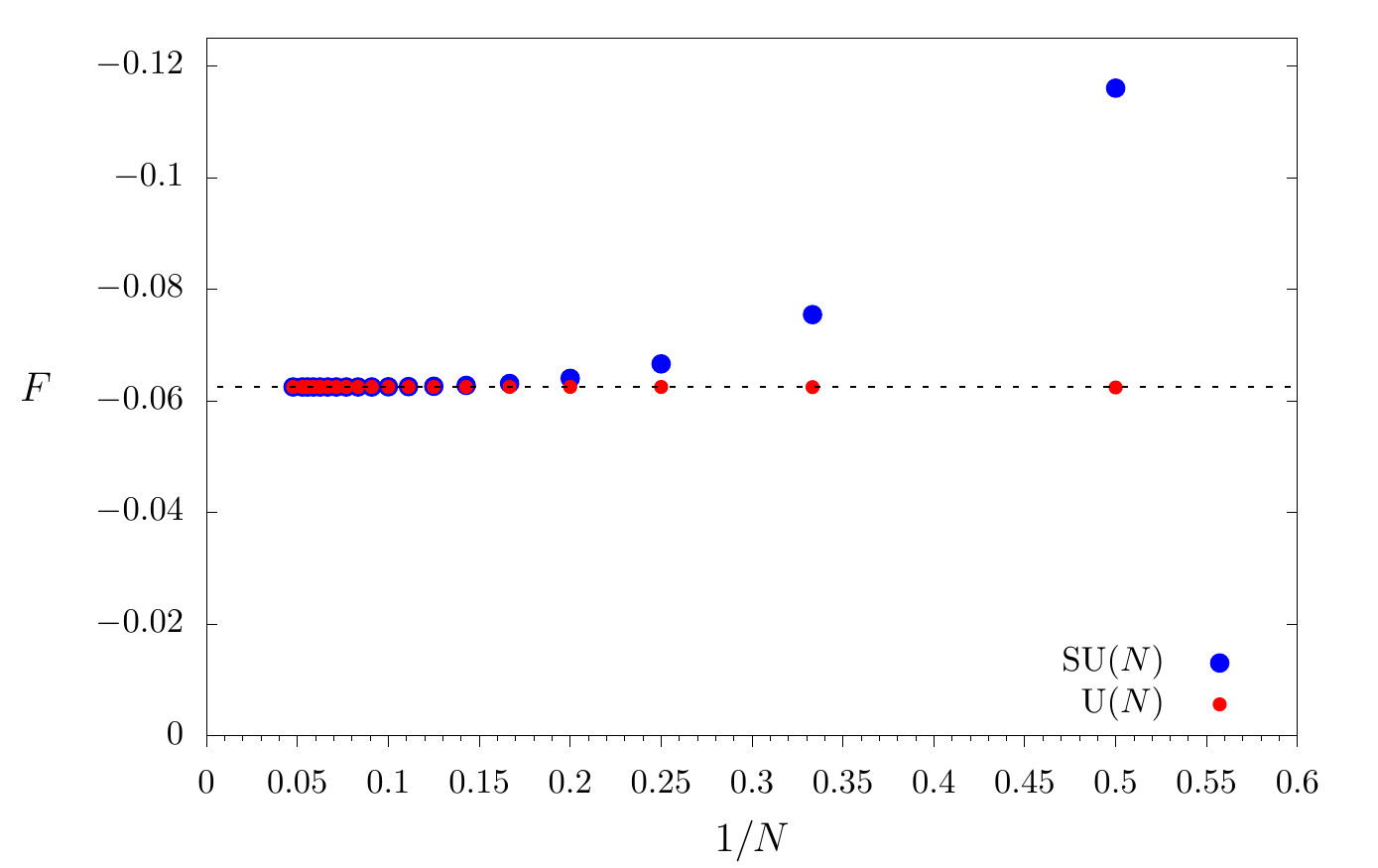}
\caption{\label{fig:plot5}The dependence of the free energy (normalized by $N^2$) 
on $N$ for $\lambda=4$ (strong coupling). There are no $1/N$ corrections for the 
\un  model while for \sun  the genus expansion does not
terminate at genus-zero.}
\end{figure}
We then consider strong coupling ($\lambda > 2$). This analysis is more
interesting since the genus expansion terminates at $\textit{genus-zero}$ in 
case of \un , first discussed in \cite{Goldschmidt:1979hq}. Our results shown 
in Figure \ref{fig:plot5} are in complete agreement with this and we note that there 
are no $1/N$ corrections in this ungapped phase
when the \sun  constraint is not imposed. For \sun  model, our study 
of the partition function signals that there are corrections and the genus 
convergence is subtle (and certainly not genus-0 exact up to instanton effects) 
in this case and this deserves further study.  

\begin{figure}[htbp] 
\centering 
\includegraphics[width=0.75\textwidth]{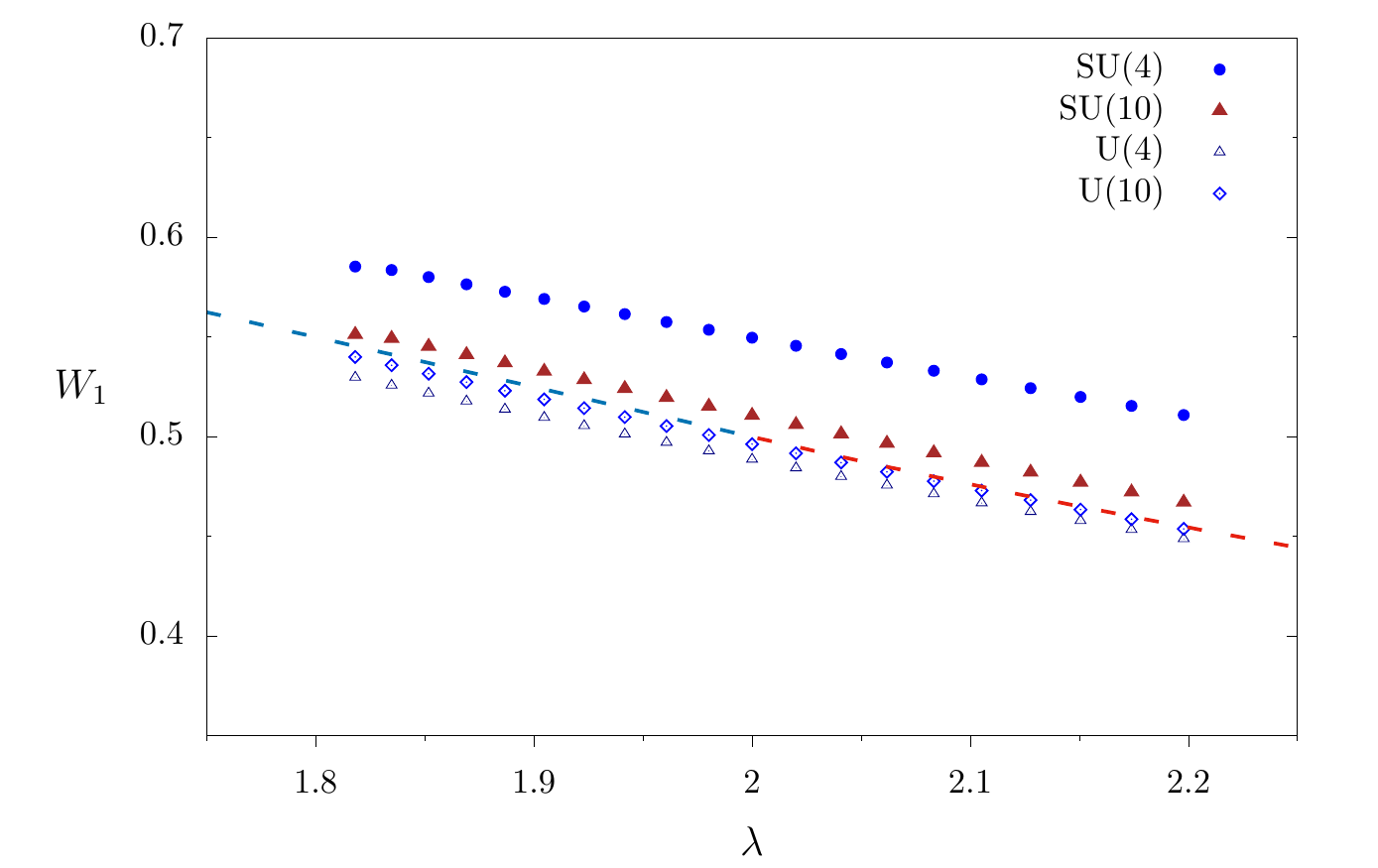}
\caption{\label{fig:plot33}The expectation value of Wilson loop against coupling for $N=4, 10$
around the critical coupling. The dashed lines (different colours) are the planar limit result in different phases.}
\end{figure}

Finally, in Figure \ref{fig:plot33}, we study the Wilson loop by taking the numerical derivative 
of the free energy for a range of couplings with $N=4, 10$ for both \sun  and \un  groups. 
The most distinct feature is that the planar result is approached from different 
sides for \sun  and \un  models (similar to free energy behaviour) and signifies
that the $1/N$ corrections to planar values come with opposite sign. It will 
be interesting to compute the exact expression for the Wilson loop winding 
around $k$ times as done for the \un  model in 
\cite{Okuyama:2017pil}. 

A related model to the one we studied here is the `double trace model' for which the action is 
$\mathrm{Tr} U \mathrm{Tr}{U}^\dagger$ and can be written in terms of the partition function of 
GWW model. This model is closely related to a 
truncated limit of $\mathcal{N}$ = 4 SYM and in the double scaling limit 
exhibits the Hagedorn/deconfinement phase transition. It would be interesting 
to understand the finite $N$ limit of this model while not restricting to 
\un  integral as done in \cite{Liu:2004vy}. 

In this paper, we have given an exact expression for the partition function of \sun one-plaquette matrix model valid for all $N$ and couplings and computed exact results for free energy and Wilson loop at finite $N$ for several couplings. 
We concluded that the $1/N$ corrections to free energy 
vanish for \un  matrix model in the ungapped (strongly coupled) phase where the only corrections to the planar result come from the instanton correction, while for \sun  matrix model, the genus expansion contribution to the free energy does not terminate at genus-zero. Our results suggest that 
that the finite $N$ behaviour of \sun  is \emph{very} different from the \un  matrix model and deserves further analysis. It would be interesting to understand results for multiply winded Wilson loops and 1/$N$ expansion of the free energy for \sun 
along the lines as done for \un  group.

\subsection*{Acknowledgements}
Research at Perimeter Institute is supported in part by the Government of Canada through the 
Department of Innovation, Science and Economic Development Canada and by the 
Province of Ontario through the Ministry of Economic Development, Job Creation and 
Trade.
\bibliographystyle{utphys}
\bibliography{oneplaq.bib}
\end{document}